\documentclass{PoS}
\usepackage{amsmath}

\title{NNLO contributions to $\epsilon_K$ and rare kaon decays}

\ShortTitle{NNLO contributions to $\epsilon_K$ and rare kaon decays}

\author{
\speaker{Martin Gorbahn} 
\thanks{
  Martin Gorbahn enjoys the hospitality of 
  the Excellence Cluster ``Origin and Structure of the Universe.''}
\\
  Technische Universit\"at M\"unchen, 
  Institute for Advanced Study, \\
  Arcisstra\ss e 21, D-80333 M\"unchen, Germany\\
  E-mail: \email{martin.gorbahn@ph.tum.de}}

\abstract{We discuss the theory prediction of $\epsilon_K$ and the
  rare $K \to \pi \nu \bar \nu$ decays and review the structure and
  current status of higher-order contributions to these flavour
  changing processes in the standard model in some detail. This
  includes the next-to-next-to-leading order QCD calculation to the
  charm quark contribution to $K^+ \to \pi^+ \nu \bar \nu$ and to the
  charm-top quark contribution to $\epsilon_K$. Electroweak
  corrections to the rare kaon decays are also discussed.}

\FullConference{2009 KAON International Conference KAON09,\\
		 June 09 - 12 2009\\
		 Tsukuba, Japan}

\begin{document}

\section{Introduction}
\label{sec:intro}

Rare decays of K-mesons as well as $K^0$-$\overline{K^0}$ mixing
continue to play an important role in fixing parameters of the
standard model (SM) and in constraining models of new physics.  In the
future $\epsilon_K$, the parameter describing indirect CP violation in
kaon mixing, and the decays $K^+ \to \pi^+ \nu \bar \nu$, $K_L \to
\pi^0 \nu \bar \nu$ will provide a decisive test of the SM and its
extensions: they are highly sensitive to new physics \cite{paride} and
their theory prediction is under good control for $\epsilon_K$ and
remarkably clean for $K \to \pi \nu \bar \nu$. In the SM these modes
are dominated by internal top quark contributions proportional to
powers of $V_{ts}^* V_{td}$ and as such are suppressed with respect to
generic new physics scenarios by the near diagonality of the
Cabibbo-Kobayashi-Maskawa (CKM) matrix. Also these modes can be
calculated precisely using an effective theory framework. The matrix
elements are extracted from $K_{l3}$ decays for $K \to \pi \nu \bar
\nu$ \cite{smith}, and from the lattice for $\epsilon_K$
\cite{boyle}. This leads to an exceptionally clean prediction for the
rare $K \to \pi \nu \bar \nu$ decays, while the recent and expected
progress for the lattice calculations of $\hat B_K$, the bag parameter
for $\epsilon_K$, is quite remarkable.

\section{Structure of $K \to \pi \nu \bar \nu$ at NLO}
\label{sec:structure}

\begin{figure}
  \begin{center}
    \includegraphics[scale=0.8]{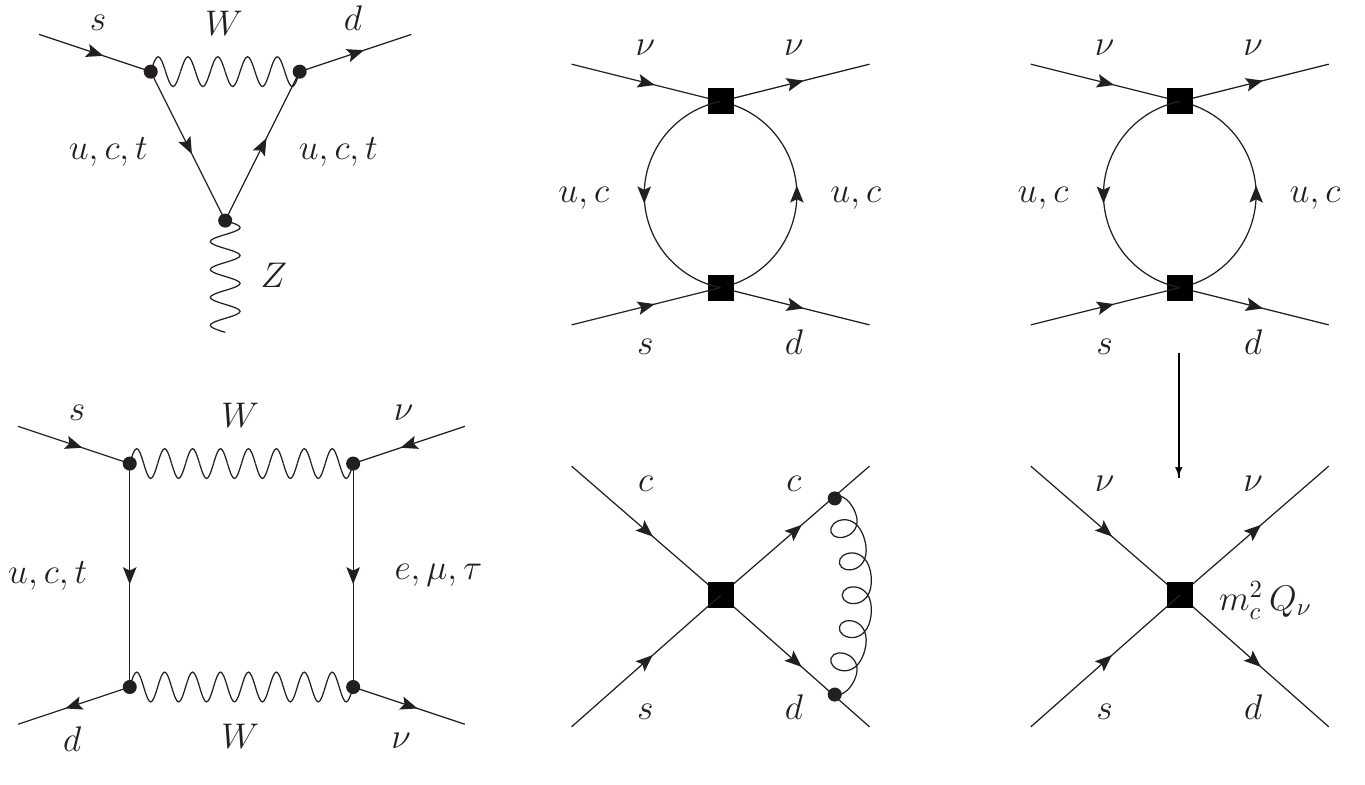}
  \end{center}
\caption{Left column: The $Z$-penguin and electroweak box give the matching
  contribution to the charm and top sector. Middle column: The mixing of
  the current-current operators into $m_c^2 \, Q_\nu$ and the
  self-mixing of the current-current operators. Right column:
  Integrating out the charm quark produces $m_c^2 Q_\nu$ and subleading
  higher dimensional operators.}
\label{fig:leading}
\end{figure}

The theoretical cleanness of the $K \to \pi \nu \bar \nu$ decays in
the SM is related to the quadratic Glashow-Iliopoulos-Maiani (GIM) 
mechanism. Using $\lambda_i = V_{is}^* V_{id}$ and $x_i = m_i^2/M_W^2$
we can write the amplitude of the $Z$-penguin and electroweak box
diagrams (Fig.~\ref{fig:leading}) as
\begin{equation}
  \label{eq:amplitude}
  \begin{split}
    \lambda_t \left( F(x_t) - F(x_u) \right) &+
    \lambda_c \left( F(x_c) - F(x_u) \right) =\\
    \mathcal{O} \left(\lambda^5 \frac{m_t^2}{M_W^2}\right) \qquad &\;+
    \left[ 
      \mathcal{O} \left( \lambda \frac{m_c^2}{M_W^2} \right) 
      \ln \frac{m_c}{M_W} +
      \mathcal{O} \left( 
        \lambda \frac{\Lambda_\textrm{QCD}^2}{M_W^2} \right)
      \ln \frac{m_c}{M_W} 
      \right].
  \end{split}
\end{equation}
Here $\lambda_t F(x_t)$ is the top quark contribution, which is
suppressed by five powers of the Cabibbo angle $\lambda = \left|
  V_\textrm{us} \right|$, and $\lambda_c F(x_c)$ is the charm quark
contribution. The contribution of soft internal up quarks is
suppressed by $\Lambda_\textrm{QCD}^2/M_W^2$.

Related to the quadratic GIM mechanism is the fact that the low-energy
effective Hamiltonian 
\begin{equation}
\label{eq:HeffSMKP}
\mathcal{H}_{\text{eff}} = \frac{4G_F}{\sqrt{2}}
\frac{\alpha}{2\pi\sin^2\theta_W} \sum_{l=e,\mu ,\tau} \left(
  \lambda_c X^l(x_c) + \lambda_t X(x_t) \right) (\bar{s}_L \gamma_{\mu}
d_L) (\bar{\nu_l}_L \gamma^{\mu} {\nu_l}_L)
\end{equation}
involves only one single operator $Q_\nu = (\bar s_L \gamma^\mu d_L)
(\bar \nu_L \gamma_\mu \nu_L)$.
The hadronic matrix element of the low-energy effective Hamiltonian can
be extracted from the well-measured $K_{l3}$ decays, including isospin
breaking and long-distance QED radiative corrections.

The anomalous dimension of the operator in \eqref{eq:HeffSMKP}
vanishes.
Hence the top quark contribution in \eqref{eq:amplitude} has no large
logarithm and is calculated in fixed-order perturbation theory. The
matching of the electroweak box and $Z$-penguin diagrams in
Fig.~\ref{fig:leading} with internal charm quarks gives the charm
quark contribution to the Wilson coefficient of $Q_\nu$. Next, one
matches Green's functions with internal $W$- and $Z$-Bosons and
dimension-six current-current operators. The bilocal mixing into the
dimension-eight operator $m_c^2 \, Q_\nu$ -- see
Fig.~\ref{fig:leading} -- resums the large logarithm in
\eqref{eq:amplitude}. The GIM mechanism cancels all loop contributions
which do not carry an explicit charm mass dependence.  Only when
integrating out the charm quark higher-dimensional operators
appear. The matching onto $Q_\nu$ in Fig.~\ref{fig:leading} gives the
dominant contribution to the branching ratio, while the contribution
of the higher-dimensional operators can be computed together with the
matrix element containing soft up quarks with the help of chiral
perturbation theory ($\chi$PT) \cite{Isidori:2005xm}.

After extracting the matrix element of $Q_\nu$ from $K_{l3}$ decays 
and summation over the three neutrino flavours the resulting branching
ratio for $K \to \pi \nu \bar \nu$ can be written as
\cite{Isidori:2005xm,Buras:2005gr,Mescia:2007kn}
\begin{equation}
\begin{split} 
  \label{eq:BR} 
  {\cal B} ( K^+ \to \pi^+ \nu \bar \nu ) &= 
  \kappa_+ \left( 1+ \Delta_\textrm{EM} \right) \left[ 
  \left ( \frac{\textrm{Im} \lambda_t}{\lambda^5} X(x_t) \right )^2 + 
  \left ( \frac{\textrm{Re} \lambda_t}{\lambda^5} X(x_t) +
    \frac{\textrm{Re} \lambda_c}{\lambda} \left (P_c + \delta P_{c,u} \right )
  \right )^2  \right] \, , \\
  {\cal B} ( K_L \to \pi^0 \nu \bar \nu ) &= \kappa_L \left (
    \frac{\textrm{Im} \lambda_t}{\lambda^5} X(x_t) \right )^2 \, .
\end{split}
\end{equation}
Here $\kappa_+ = 0.5173(25)\times 10^{-10}(\lambda/0.225)^8$
and $\kappa_L= 2.231(13)\times 10^{-10}(\lambda/0.225)^8$
contain higher-order electroweak corrections for the
normalisation to the $K_{l3}$ decays, and $\Delta_{\text{EM}} \simeq
-0.3\%$ denotes long distance QED corrections \cite{Mescia:2007kn}.
The top quark $X(x_t)=1.464 \pm 0.041$, computed at two-loop in
Ref.~\cite{Buchalla:1993wq}, gives the only contribution to ${\cal
  B}(K_L \to \pi^0 \nu \bar \nu)$, while its contribution to ${\cal
  B}(K^+ \to \pi^+ \nu \bar \nu)$ is 63\%. The perturbative part of
the charm quark contribution at NLO \cite{Buchalla:1993wq} is
\begin{equation}
  \label{eq:1}
  P_c^{\textnormal{NLO}} = 0.364 \pm 0.036_{\rm theory} \pm 
  0.009_{m_c} \pm 0.009_{\alpha_s} \, ,
\end{equation}
for $\lambda = 0.2255$. The parametric uncertainty -- see
Ref.~\cite{Brod:2008ss} for input parameters -- is small compared to
the theoretical error, which results from higher order corrections. In
a $\chi$PT calculation \cite{Isidori:2005xm} the contribution of
higher dimensional operators and soft up quarks has been calculated
to $\delta P_{c,u}=0.04 \pm 0.02$.  Using Eq.~\eqref{eq:BR},
Eq.~\eqref{eq:1}, and the input parameters of Ref.~\cite{Brod:2008ss}
results in:
\begin{equation}
  \label{eq:5}
  \begin{split}
   {\cal B}(K^+ \to \pi^+ \nu \bar \nu)^\textrm{NLO} &= ( 8.5 
  \pm 0.5_{\rm sd} 
  \pm 0.2_{\rm ld}
  \pm 0.6_{\rm param}) \times 10^{-11} \, ,\\
   {\cal B}(K_L \to \pi^0 \nu \bar \nu)^\textrm{NLO} &= 
  ( 2.7   
  \pm 0.1_{\rm sd} 
  \pm 0.04_{\rm ld}
  \pm 0.4_{\rm param} ) \times 10^{-11} \, .
  \end{split}
\end{equation}
The subscript ``sd'', ``ld'', and ``param'' labels the perturbative,
long-distance, and parametric uncertainties respectively. The
parametric error is dominated by the CKM parameters, while $\delta
P_{c,u}$ gives the largest contribution to the long-distance
uncertainty of the charged decay mode.

\section{Structure of $\epsilon_K$ at NLO}

\label{sec:eK}
The parameter 
\begin{equation}
  \label{eq:ekexpdef}
  \epsilon_K = \frac{
    \mathcal{A} \left ( K_L \to (\pi \pi)_{I=0} \right )}{
    \mathcal{A} \left ( K_S \to (\pi \pi)_{I=0} \right )}
\end{equation}
measures CP violation in $K_0$--$\overline{K_0}$ mixing via the ratio
of the respective decay amplitudes of a $K_L$ and a $K_S$ decaying
into a two pion state of isospin zero. Its generic structure shares
features with the previously discussed $K \to \pi \nu \bar \nu$
decays: only one operator $Q_{S2} = (\bar s \gamma_\mu d_L)(\bar s
\gamma_\mu d_L)$ contributes dominantly below the charm quark mass
scale and long-distance effects are power suppressed, this time
because of CP violation.

For the theoretical prediction it is useful to express $\epsilon_K$ in
terms of $\langle K^0 | \mathcal{H}_{eff}^{\Delta S=2} | \bar K^0
\rangle = 2 m_K M_{12}^*$, the matrix element of the $\Delta S =2$
effective Hamiltonian, and write:
\begin{equation}
  \label{eq:ektheo}
  \epsilon_K = 
  e^{i \phi_\epsilon} \sin \phi_\epsilon 
  \left( \frac {\textrm{Im} (M_{12}^*)}  {\Delta M_K} + \xi \right) \, .
\end{equation}
Here the phase of $\epsilon_K$ is $\phi_\epsilon = 43.5(7)^{\circ}$
\cite{Amsler:2008zzb} and $\xi = \textrm{Im} A_0/\textrm{Re} A_0
\simeq 0$ is the imaginary part divided by the real part of the
isospin zero amplitude $A_0 = \mathcal{A} \left ( K_S \to (\pi
  \pi)_{I=0} \right )$. The ratio
$\kappa_\epsilon=|\epsilon_K^{SM}/\epsilon_K(\phi_\epsilon=45^{\circ},
\xi = 0 )|$ encompasses the change of $|\epsilon_K|$ if the values
$\phi_\epsilon = 45^{\circ}$ and $\xi = 0$ are used
in~(\ref{eq:ektheo}), as has been done in most of the older analyses,
instead of the exact values. The authors of Ref.~\cite{Buras:2009pj}
used $\epsilon'/\epsilon$ to extract the value of $\kappa_\epsilon =
0.92 \pm 0.02$ in the SM.

\begin{figure}
  \begin{center}
    \begin{tabular}{cc}
      \includegraphics[scale=0.6]{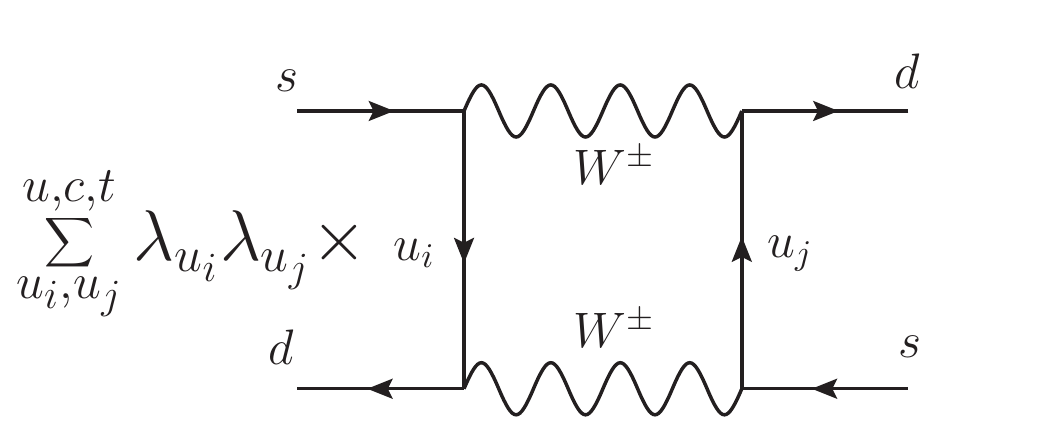} &
      \includegraphics[scale=0.6]{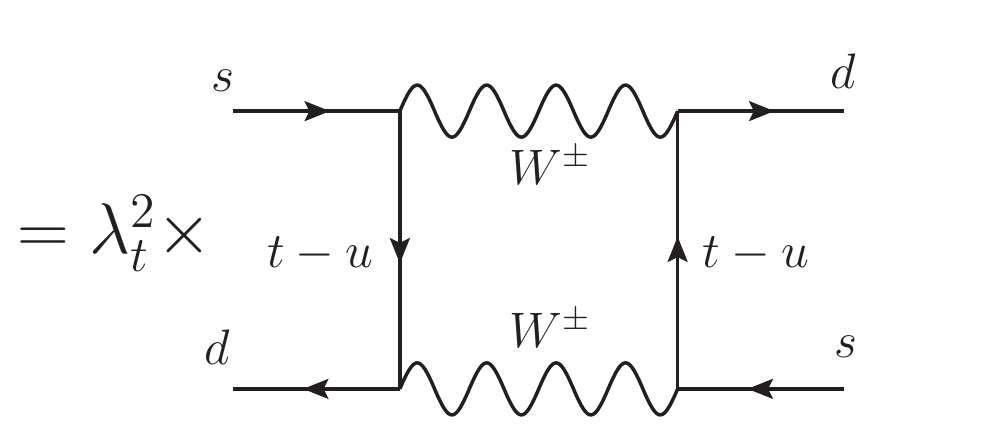} \\
      a) & b) \\
      \includegraphics[scale=0.6]{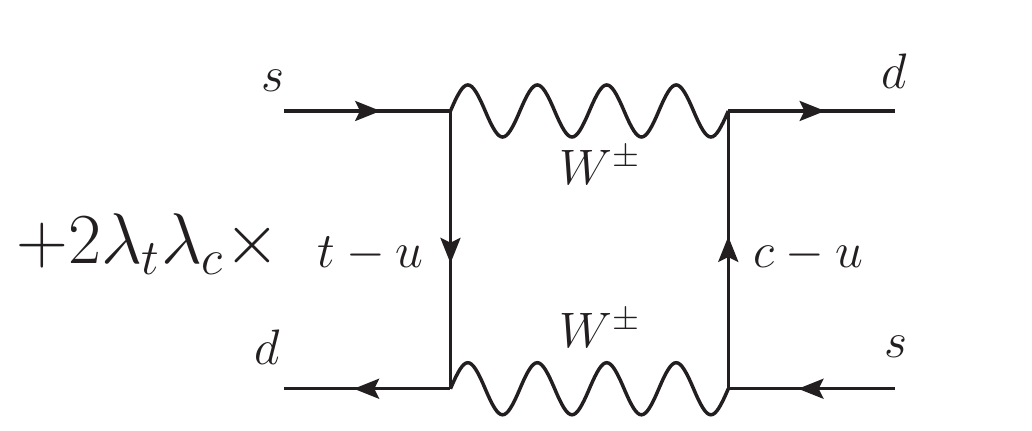} &
      \includegraphics[scale=0.6]{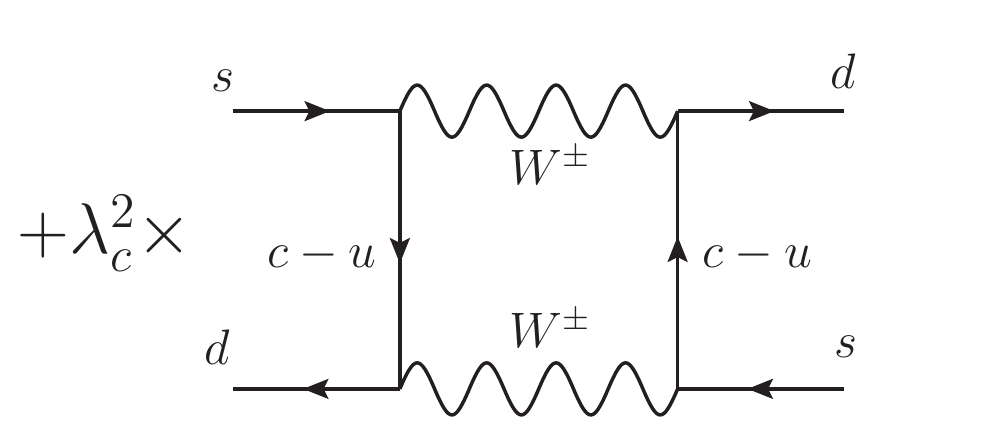} \\
      c) & d)
    \end{tabular}
  \end{center}
  \caption{ The $\Delta S=2$ Box-type diagram with internal up,
    charm, and top contributions is expressed as a sum of Box-type
    diagrams proportional to $\lambda_t^2$, $\lambda_c^2$, and
    $\lambda_t \lambda_c$ respectively using the GIM mechanism.}
  \label{fig:leadingbox}
\end{figure}
The box diagram of Fig.~\ref{fig:leadingbox}a gives the leading
contribution to the $\Delta S=2$ effective Hamiltonian and the
parameter $M_{12}$. It is proportional to a sum of loop functions
times CKM factors:
\begin{equation}
  \label{eq:GIMforS}
  \sum_{u_i,u_j \in \{ u,c,t \}}
  \lambda_{u_i} \lambda_{u_j} 
  \tilde S(\frac{m_{u_i}^2}{M_W^2},\frac{m_{u_j}^2}{M_W^2}) = 
  \lambda_t^2 S(x_t) + 
  \lambda_c^2 S(x_c) + 
  \lambda_t \lambda_c S(x_t,x_c) \; .
\end{equation}
After the GIM mechanism has been used to eliminate $\lambda_u = -
\lambda_t - \lambda_c$ it comprises the top quark contribution --
proportional to $\lambda_t^2$ (Fig.~\ref{fig:leadingbox}b), the charm
quark contribution -- proportional to $\lambda_c^2$
(Fig.~\ref{fig:leadingbox}c), and the charm-top quark contribution
(Fig.~\ref{fig:leadingbox}d) -- proportional to $\lambda_t
\lambda_c$. The resulting loop functions $S(x_i,x_j) = \tilde
S(x_i,x_j) - \tilde S(x_i,0) - \tilde S(0,x_j) + \tilde S(0,0)$ and
$S(x_i) = S(x_i,x_i)$ are suppressed by the smallness of the quark
mass $m_i$ if $x_i =m_i^2/M_W^2$ is significantly smaller than
one. This, together with the severe Cabibbo suppression of the CP
violating top quark contribution, lets all three contributions compete
in size for $\epsilon_K$:
\begin{equation}
  \label{eq:sizecon}
  \textrm{Im} \left(
    \lambda_t^2 S(x_t) + 
    \lambda_c^2 S(x_c) + 
    \lambda_t \lambda_c S(x_t,x_c)
  \right) \simeq 
  \mathcal{O}\left(\lambda^{10}\right) + 
  \mathcal{O}\left(\lambda^{6} \frac{m_c^2}{M_W^2}\right)
  \ln \left(\frac{m_c}{M_W} \right) +
  \mathcal{O}\left(\lambda^{6} \frac{m_c^2}{M_W^2}\right) \; .
\end{equation}
The diagram of Fig.~\ref{fig:leadingbox}a induces a large logarithm
($\ln m_c/M_W$) only for the charm-top quark contribution: the large
logarithm from the up-quarks in Fig.~\ref{fig:leadingbox}b is power
suppressed by $\Lambda_\textrm{QCD}^2/M_W^2$, while the GIM mechanism
cancels a potential $\ln m_c/M_W$ between the diagrams with both one
up and one charm quark and the diagram with only internal charm
quarks.

\begin{figure}
  \centering
    \begin{tabular}{cc}
      \includegraphics[scale=0.6]{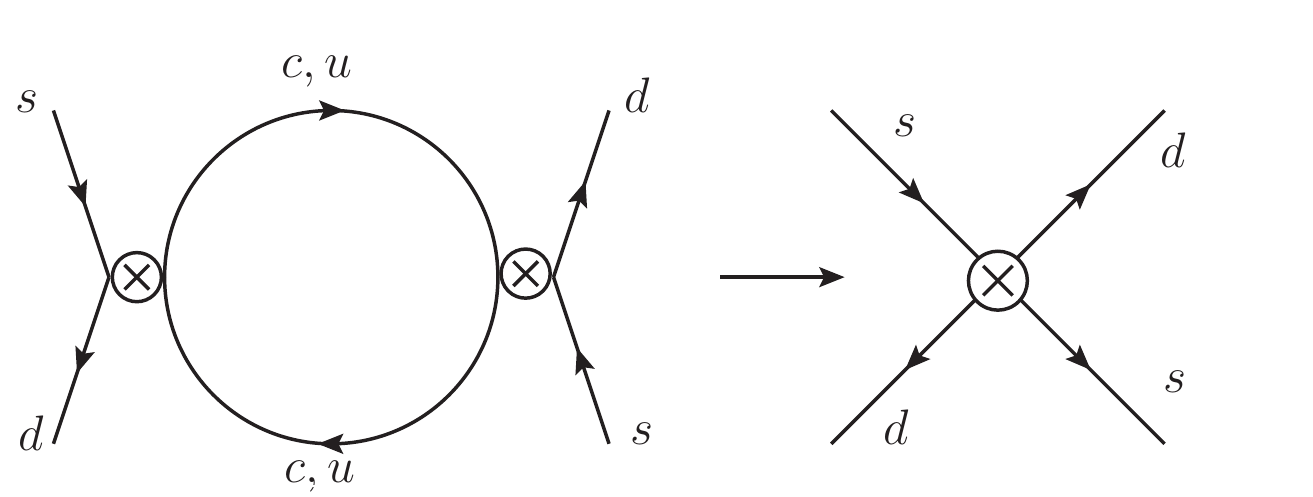} &
      \includegraphics[scale=0.6]{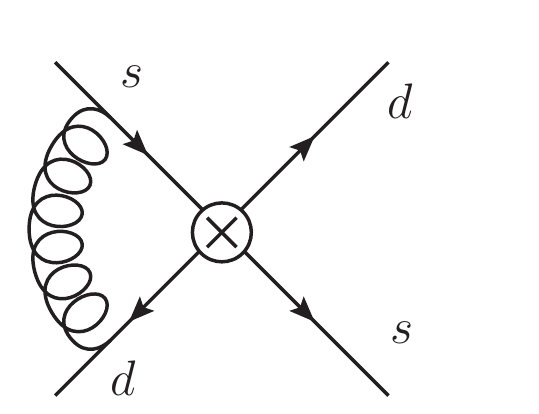} \\
      a) & b) 
    \end{tabular}
    \caption{Dimension 6 current-current and penguin operators mix at
      LO into $Q_{S2}$ with a CKM factor proportional to $\lambda_t
      \lambda_c$ in a). Integrating out the charm quark results in
      similar diagrams for the LO and NLO matching of the contribution
      proportional to $\lambda_t \lambda_c$ and $\lambda_c^2$
      respectively. A sample diagram which is relevant to the LO
      evolution of $Q_{S2}$ is shown in b).}
  \label{fig:mixinglo}
\end{figure}

This can be reformulated in an effective theory language: the
dimension-six penguin as well as the current-current operators, which
have tree-level Wilson coefficients, mix only into the charm-top quark
contribution, via the bilocal mixing in Fig.~\ref{fig:mixinglo}a, yet
do not induce large logarithms times tree-level Wilson coefficients
proportional to $\lambda_t^2$ and $\lambda_c^2$. QCD corrections do
not change this picture but only induce the well known RGE effects for
the $\Delta S=1$ effective Hamiltonian \cite{Gorbahn:2004my} and for
the $\Delta S=2$ Operator $Q_{S2}$ (Fig.~\ref{fig:mixinglo}b). A LO
analysis of the charm quark and top quark contribution to $\epsilon_k$
then requires a one-loop calculation both for the matching at $\mu_W$,
for the running, and for the matching of the charm quark contribution
also for the matching at $\mu_c$ (Fig.~\ref{fig:mixinglo}a). This is
contrary to the charm-top quark contribution where a tree-level
matching at $\mu_W$ and $\mu_c$ is sufficient at LO.

After integrating out the charm quark the $\Delta S=2$ effective
Hamiltonian reads
\begin{equation}
    \mathcal{H}^{\Delta S=2}_\textrm{eff} = \frac{G_F^2}{4 \pi^2} M_W^2 \left[
      \lambda_c^2 \eta_\textrm{cc} S(x_c) +
      \lambda_c^2 \eta_\textrm{tt} S(x_t) +
      \lambda_c^2 \eta_\textrm{ct} S(x_c,x_t) \right] b(\mu) Q_{S2} + 
    \textrm{h.c.} + \dots
  \label{eq:Hlo}
\end{equation}
where the QCD and logarithmic corrections are known at NLO and
parametrised by $\eta_{\rm cc}=1.43(23)$ \cite{Herrlich:1993yv},
$\eta_{\rm ct}=0.47(4)$ \cite{Herrlich:1993yv}, and
$\eta_{\rm tt}=0.5765(65)$ \cite{Buras:1990fn}. The parameter $b(\mu)$ is
factored out such that
\begin{equation}
  \label{eq:bkpar}
  \hat B_K = \frac{3}{2} b(\mu) 
  \frac{
    \langle K^0 | Q_{S2} | \overline{K^0}\rangle}{
    f_K^2 m_K^2}
\end{equation}
is a renormalisation group invariant quantity, which can be calculated
on the lattice -- see e.g. \cite{Antonio:2007pb} and Ref.\cite{boyle}
of this conference. Using $\hat B_K=0.720 \pm 0.013 \pm 0.037$ one
finds for $\epsilon_K$ at NLO \cite{Buras:2009pj}:
\begin{equation}
  \label{eq:epsknlo}
  \epsilon_K^\textrm{NLO} = (1.78 \pm 0.25) \; ,
\end{equation}
where $\eta_{\rm tt}$, $\eta_{\rm ct}$, and $\eta_{\rm cc}$ contribute
with 75\%, 37\%, and $-12$\% respectively to the total value of $\epsilon_K$,
while 60\% of the uncertainty is of parametric origin and 40\% is of
theoretical origin. The parametric error is dominated by the
uncertainty in the CKM parameters, while the perturbative and
non-perturbative uncertainties are comparable in size for the theory
uncertainty.

Finally note that $\mathcal{H}^{\Delta S=2}$ also contains higher
dimensional Operators in $\mathcal{H}^{\Delta S=2}_\textrm{eff,d=8}$
and current-current operators with up-quarks in $\mathcal{H}^{\Delta
  S=1}_\textrm{eff,up}$, as indicated by the ellipses in 
Eq.~(\ref{eq:Hlo}). At leading order in the $1/N_C$ expansion only one
higher-dimensional operator is present and its matrix element is
estimated in \cite{Cata:2004ti} to result in a $0.5\%$ enhancement of
$\epsilon_K$.

\section{NNLO QCD corrections}
\label{sec:nnlo}

The NNLO calculation of $\epsilon_K$ and $K^+ \to \pi^+ \nu \bar \nu$
aims at resumming all ${\cal O} (\alpha_s^n \ln^{n-1} (
\mu_W^2/\mu_c^2 ))$ logarithms for $P_c$ and $\eta_\textrm{ct}$ and
all ${\cal O} (\alpha_s^n \ln^{n-2} (\mu_W^2/\mu_c^2 ))$ for $X(x_t)$,
$\eta_\textrm{tt}$, and $\eta_\textrm{cc}$. The theory uncertainty of
$P_c$ and $\eta_\textrm{ct}$ dominates the perturbative error for $K^+
\to \pi^+ \nu \bar \nu$ and $\epsilon_K$ respectively at NLO, while
the large theory uncertainty in $\eta_\textrm{cc}$ is somewhat
suppressed by the smallness of the charm contribution to $\epsilon_K$.

A NNLO analysis for $P_c$ and $\eta_\textrm{ct}$ will reduce the
theory uncertainties and comprises $(i)$ the ${\cal O} (\alpha_s^2)$
matching corrections to the relevant Wilson coefficients arising at
$\mu_W$, $(ii)$ the ${\cal O} (\alpha_s^3)$ anomalous dimensions
describing the mixing of the dimension-six and the $Q_\nu$ and
$Q_\textrm{S2}$ operators, $(iii)$ the ${\cal O} (\alpha_s^2)$
threshold corrections to the Wilson coefficients originating at
$\mu_b$, and $(iv)$ the ${\cal O} (\alpha_s^2)$ matrix elements of the
operators emerging at $\mu_c$.
\begin{figure}
  \begin{center}
    \includegraphics[scale=1.5]{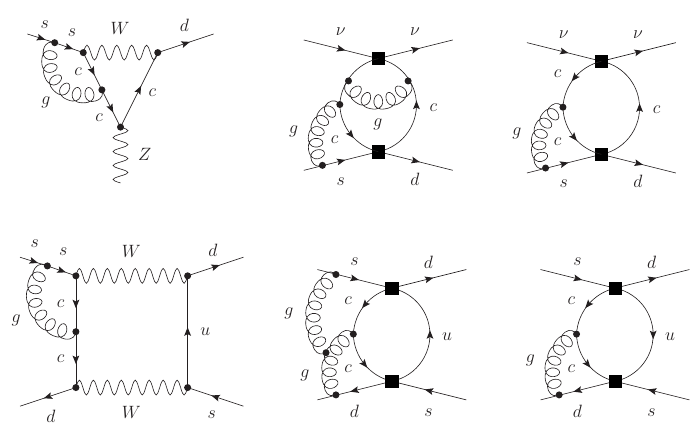}
  \end{center}
  \caption{Examples of Feynman diagrams arising in the full SM (left
    column), describing the mixing of operators (centre column) and
    the matrix elements (right column) for the $Z$-penguin sector of
    $K^+ \to \pi^+ \bar \nu \nu$ (upper row) and for $\epsilon_K$
    (lower row). Only the divergent pieces of the diagrams displayed
    in the centre column have to be computed, while the Feynman graphs
    shown on the left- and right-hand side are needed including their
    finite parts.}
\label{fig:diagrams}
\end{figure}
To determine the contributions of type $(i)$, $(iii)$, and $(iv)$ one
must calculate two-loop Green functions in the full SM and in
effective theories with five or four flavours. Sample diagrams for
steps $(i)$ and $(iv)$ are shown in the left and right columns of
Fig.\,\ref{fig:diagrams}.  The contributions $(ii)$ are found by
calculating three-loop Green functions with operator
insertions. Sample diagrams with a double insertion of dimension-six
operators are shown in the centre column of
Fig.\,\ref{fig:diagrams}. The corresponding three-loop amplitudes are
evaluated using the method that has been described in
\cite{Gorbahn:2004my,ADM}. A comprehensive discussion of the technical
details of the matching, the renormalisation of the effective theory
and the actual calculation is given in \cite{Buras:2005gr} for the
calculation of $P_c$ and will be given in \cite{eknnlo} for the
calculation of $\eta_\textrm{ct}$. The same techniques have also been
used to reduce the uncertainties in the short-distance contribution to
$K_L \to \mu^+ \mu^-$ \cite{Gorbahn:2006bm}.

The aforementioned QCD calculation results for the input parameters of
Ref.~\cite{Brod:2008ss} in the following value for $P_c$ at NNLO:
\begin{equation}
  \label{eq:3}
  P_c^{\textrm{NNLO}}  = 0.368 \pm 0.009_{\rm theory} \pm 0.009_{m_c} \pm 0.009_{\alpha_s}
  \, .
\end{equation}
Comparing these numbers with Eq.~\eqref{eq:1} we observe that the NNLO
calculation reduces the theoretical uncertainty by a factor of
$4$. Because of this and the improvement in the calculation of
long-distance effects \cite{smith}, unknown electroweak corrections
could potentially dominate the theory uncertainty of the rare $K \to
\pi \nu \bar \nu$ decays. Even though a similar reduction of the error
for $\eta_\textrm{ct}$ is expected at NNLO in QCD \cite{eknnlo} no
electroweak corrections are needed for the present theoretical status
of $\epsilon_K$.

\section{Electroweak corrections}
\label{sec:ewcor}

The NLO calculation of electroweak corrections for rare $K \to \pi
\bar \nu \nu$ decays resums all LO and NLO logarithmic QED corrections
and fixes the scheme electroweak input parameters, like $\sin^2
\theta_W$, by an electroweak matching calculation. The function $P_c$
depends on the charm quark $\overline{\textrm{MS}}$ mass through the
parameter $x_c$, conventionally defined as $x_c = m_c^2/M_W^2$.
The point of fixing the input parameters can be exemplified by noting
that the charm quark contribution is mediated by a double insertion of
two dimension-six operators. This results in a contribution of
$\mathcal{O} (G_F^2)$ -- the second power of $G_F$ resides in $x_c$ --
plus electroweak corrections. Yet the leading result of
Eq.~(\ref{eq:HeffSMKP}) can only approximate the electroweak
corrections for a specific choice of the renormalisation scheme for
the prefactor of the charm quark contribution, expressed as
$\alpha/\sin^2 \theta_W$. While it is expected that using
$\overline{\textrm{MS}}$ parameters renormalised at the electroweak
scale would approximate the electroweak corrections best
\cite{Bobeth:2003at}, only an explicit calculation can provide a
definite result. We normalise all dimension-six operators
to $G_F$ and replace the parameter $x_c$ 
with the definition
\begin{equation}
  \label{eq:xc}
  x_c = \sqrt{2} \frac{\sin^2 \theta_W}{\pi \alpha} G_F m_c^2 (\mu_c) \, ,
\end{equation}
which only at tree level equals the ratio $m_c^2 (\mu_c)/M_W^2$.

\begin{figure}
  \begin{minipage}[Anordnung]{0.33 \linewidth}
    \includegraphics[scale=0.88]{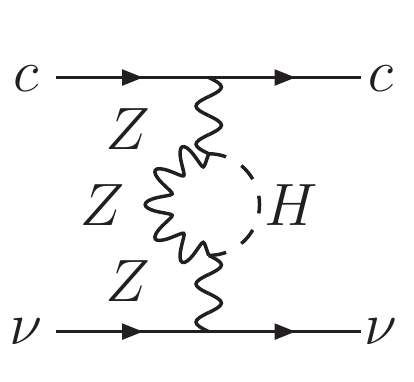} \\[2mm]
    \includegraphics[scale=0.55]{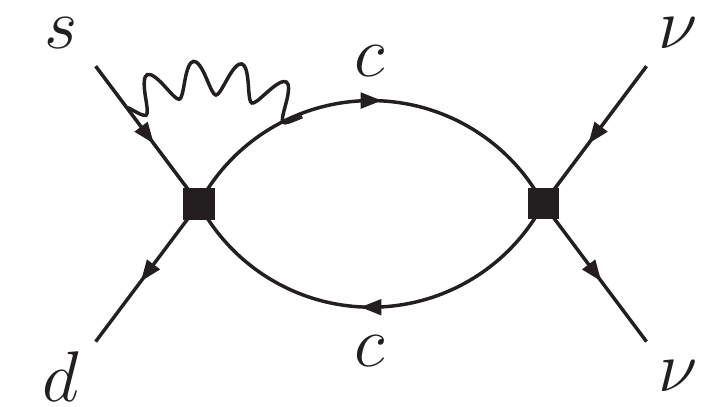}\\[0mm]
  \end{minipage}
  \begin{minipage}[Anordnung]{0.5 \linewidth}
    \includegraphics[scale=0.8]{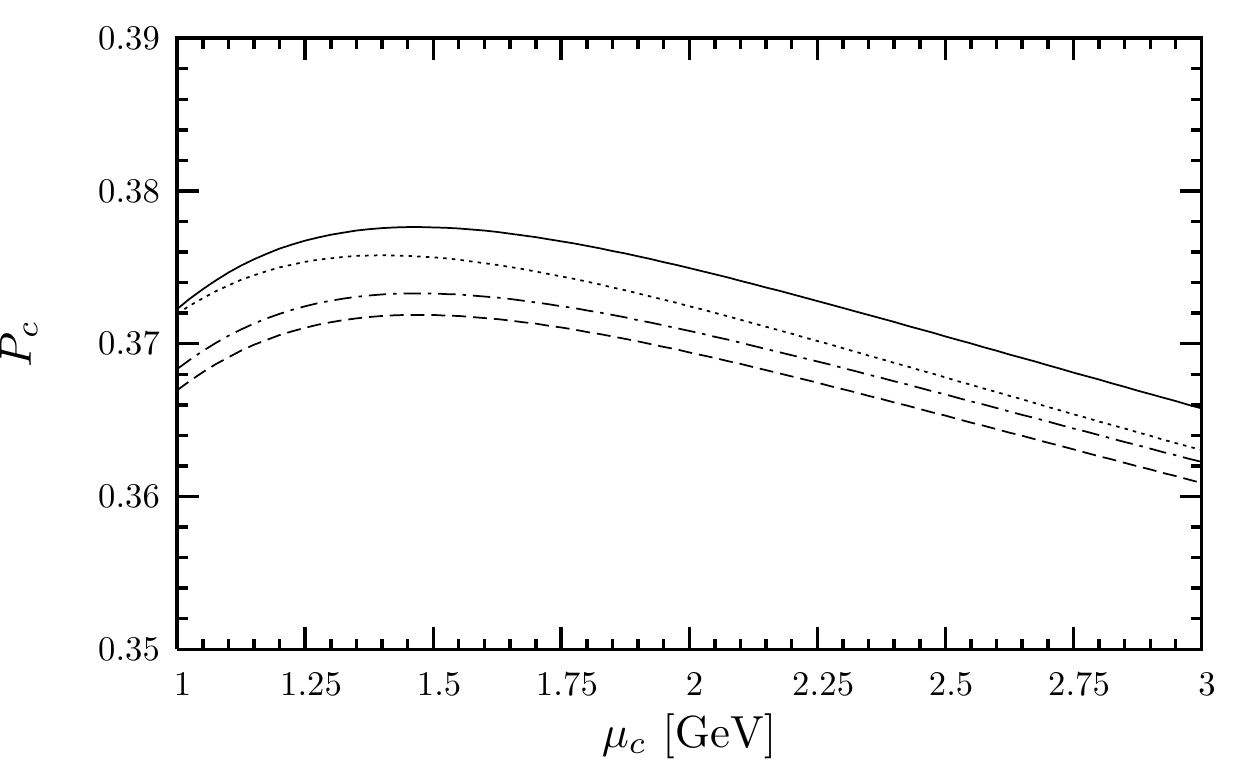}
  \end{minipage}
  \caption{Left column: Example of a diagram describing the NLO
    matching to a dimension-six operator involving charm quarks and
    neutrinos (top) and of a diagram contributing to the NLO mixing of
    two dimension-six operators into $Q_\nu$. Right column: $P_c(X)$
    as a function of $\mu_c$ at NNLO QCD (dashed dotted line),
    including LO QED (dotted line), and NLO electroweak corrections
    (solid line). The dashed line shows $P_c(X)$ at NNLO QCD where the
    definition $x_c=m_c^2/M_W^2$ is used.}
\label{fig:plots}
\end{figure}

The NLO analysis of electroweak effects of Ref.~\cite{Brod:2008ss}
involves the calculation of one-loop matching corrections for the
dimensions-six operators (top left diagram of Fig.~\ref{fig:plots})
and QED corrections to the LO QCD operator mixing (bottom left diagram
of Fig.~\ref{fig:plots}), and the inclusion of QED effects in the
expansion of the matrix elements at $\mu_c$. 
Note that the LO QED corrections start at $\mathcal{O} (\alpha^2 \ln^2
(\mu_W^2/\mu_c^2 ))$ while the first NLO electroweak correction is
$\mathcal{O} (\alpha^2 \ln (\mu_W^2/\mu_c^2 ))$. This explains why
$P_c(x)$, which is plotted on the right column of Fig.~\ref{fig:plots}
as a function of the parameter $\mu_c$, receives corrections of
similar size. Also the cancellation of the scheme dependence between
the LO QED and the NLO electroweak contribution is clearly visible and
we see that including the full electroweak corrections, $P_c(X)$ is
mildly increased as compared to the pure NNLO QCD. The 
number for the branching ratio then reads \cite{Brod:2008ss}:
\begin{equation}\label{eq:BRnum}
  \mathcal{B}\left(K^+\to\pi^+\nu\bar{\nu}\right) =
  (8.51^{\, +0.57}_{\, -0.62\; _\textrm{CKM}} \pm 
  0.20_{m_c,m_t,\alpha_s} \pm 
  0.36_{\textrm{theory}})\times 10^{-11} \; . 
\end{equation}
The CKM parameters dominate the parametric uncertainty. The main
contributions to the theory error stem from the uncertainty in $\delta
P_{c,u}$ and $X_t$, where we used an error of $2\%$. In detail, the
contributions to the theory error are ($\kappa_{\nu}^+: 6\%$, $X_t:
38\%$, $P_c: 17\%$, $\delta P_{c,u}: 39\%$), respectively. All errors
have been added in quadrature.

\section{Conclusions}

Rare $K \to \pi \nu \bar \nu$ decays and the CP violating parameter
$\epsilon_K$ are extremely sensitive to flavour violating new
physics. The good control of long-distance contribution to these
observables makes the calculation of NNLO QCD and sometimes even NLO
electroweak corrections mandatory. Results for NNLO QCD and NLO
electroweak corrections for the charm quark contributions to rare K
decays have been published in Ref.~\cite{Buras:2005gr,Brod:2008ss},
while the NNLO calculation of the charm-top quark contribution is
finished \cite{eknnlo}. This, together with current \cite{Ahn:2006uf}
and future \cite{exp} progress from the experimental side, will
increase the new physics reach of these observables further.

\acknowledgments{I would like to thank the organisers of the KAON09
conference for the invitation to such an interesting meeting. A big
thank you to J. Brod for his careful reading of this manuscript.}


\begin{thebibliography}{99}

\bibitem{paride} 
P.~Paradisi, \pos{PoS(KAON09)044}.

\bibitem{smith} 
C.~Smith, \pos{PoS(KAON09)010}.

\bibitem{boyle}
 P.~Boyle, \pos{PoS(KAON09)002}. 

\bibitem{Isidori:2005xm}
G.~Isidori, F.~Mescia and C.~Smith,
Nucl.\ Phys.\ B {\bf 718}, 319 (2005).

\bibitem{Buras:2005gr}
A.~J.~Buras, M.~Gorbahn, U.~Haisch, and U.~Nierste,
Phys.\ Rev.\ Lett.\ {\bf 95}, 261805 (2005);
A.~J.~Buras, M.~Gorbahn, U.~Haisch, and U.~Nierste,
[arXiv:hep-ph/0603079].

\bibitem{Mescia:2007kn}
  F.~Mescia and C.~Smith,
  Phys.\ Rev.\  D {\bf 76} (2007) 034017
  [arXiv:0705.2025 [hep-ph]].

\bibitem{Buchalla:1993wq}
G.~Buchalla and A.~J.~Buras,
Nucl.\ Phys.\ B {\bf 412}, 106 (1994),
  M.~Misiak and J.~Urban,
  Phys.\ Lett.\  B {\bf 451} (1999) 161
  [arXiv:hep-ph/9901278],
G.~Buchalla and A.~J.~Buras,
Nucl.\ Phys.\ B {\bf 548}, 309 (1999);

\bibitem{Brod:2008ss}
  J.~Brod and M.~Gorbahn,
  Phys.\ Rev.\  D {\bf 78} (2008) 034006
  [arXiv:0805.4119 [hep-ph]].

\bibitem{Amsler:2008zzb}
  C.~Amsler {\it et al.}  [Particle Data Group],
  Phys.\ Lett.\  B {\bf 667} (2008) 1.

\bibitem{Buras:2009pj}
  A.~J.~Buras and D.~Guadagnoli,
  Phys.\ Rev.\  D {\bf 79} (2009) 053010
  [arXiv:0901.2056 [hep-ph]].

\bibitem{Gorbahn:2004my}
M.~Gorbahn and U.~Haisch,
Nucl.\ Phys.\ B {\bf 713}, 291 (2005). 

\bibitem{Herrlich:1993yv}
  S.~Herrlich and U.~Nierste,
  Nucl.\ Phys.\  B {\bf 419} (1994) 292
  [arXiv:hep-ph/9310311];
  S.~Herrlich and U.~Nierste,
  Phys.\ Rev.\  D {\bf 52} (1995) 6505
  [arXiv:hep-ph/9507262];
  S.~Herrlich and U.~Nierste,
  Nucl.\ Phys.\  B {\bf 476} (1996) 27
  [arXiv:hep-ph/9604330].

\bibitem{Buras:1990fn}
  A.~J.~Buras, M.~Jamin and P.~H.~Weisz,
  Nucl.\ Phys.\  B {\bf 347}, 491 (1990).

\bibitem{Antonio:2007pb}
  D.~J.~Antonio {\it et al.}  [RBC Collaboration and UKQCD Collaboration],
  Phys.\ Rev.\ Lett.\  {\bf 100} (2008) 032001
  [arXiv:hep-ph/0702042];
  C.~Allton {\it et al.}  [RBC-UKQCD Collaboration],
  Phys.\ Rev.\  D {\bf 78} (2008) 114509
  [arXiv:0804.0473 [hep-lat]];
  C.~Aubin, J.~Laiho and R.~S.~Van de Water,
  arXiv:0905.3947 [hep-lat].

\bibitem{Cata:2004ti}
  O.~Cata and S.~Peris,
  JHEP {\bf 0407} (2004) 079
  [arXiv:hep-ph/0406094].

\bibitem{ADM}
K.~G.~Chetyrkin, M.~Misiak and M.~M\"unz,
Nucl.\ Phys.\ B {\bf 518}, 473 (1998);
P.~Gambino, M.~Gorbahn and U.~Haisch,
Nucl.\ Phys.\ B {\bf 673}, 238 (2003).

\bibitem{eknnlo} 
  J.~Brod and M.~Gorbahn, in preparation; 
  J.~Brod,
  PhD Thesis, Karlsruhe, 2009; 

\bibitem{Gorbahn:2006bm}
  M.~Gorbahn and U.~Haisch,
  Phys.\ Rev.\ Lett.\  {\bf 97} (2006) 122002
  [arXiv:hep-ph/0605203].

\bibitem{Bobeth:2003at}
  C.~Bobeth, P.~Gambino, M.~Gorbahn and U.~Haisch,
  JHEP {\bf 0404} (2004) 071
  [arXiv:hep-ph/0312090].

\bibitem{Ahn:2006uf}
  J.~K.~Ahn  [E391a Collaboration],
  arXiv:hep-ex/0607016.

\bibitem{exp}
 G. Ruggiero, \pos{PoS(KAON09)043}; 
 H. Nanjo, \pos{PoS(KAON09)047}; 
 D. A. Bryman, \pos{PoS(KAON09)049}. 

\end{thebibliography}
\end{document}